\documentclass[amsmath,amssymb,aps,prb,twocolumn,fleqn]{revtex4}

\newcommand{\ket}[1]{\ensuremath{|#1\rangle}}

\newcommand{\Jp}{\ensuremath{J^{\prime}}}
\newcommand{\approxgt}{ \,{\scriptstyle\gtrsim} \,}
\newcommand{\approxlt}{ \,{\scriptstyle\lesssim} \,}
\newcommand{\heff}{{H}_{\mbox{\tiny{eff}}}}
\newcommand{\mch}{{H}}
\newcommand{\mchain}{m_{\mbox{\tiny ch}}}

\usepackage{graphics}

\begin{document}

\title{Ordered Phases of the Anisotropic Kagome Lattice Antiferromagnet in a Field}
\date{\today}
\author{E.M. Stoudenmire}
\affiliation{Department of Physics, University of California, Santa Barbara, CA 93106-9530}
\author{Leon Balents}
\affiliation{Department of Physics, University of California, Santa Barbara, CA 93106-9530}

\begin{abstract}
  The antiferromagnetic Heisenberg model on an anisotropic kagome
  lattice may be a good minimal model for real magnetic systems as
  well as a limit from which the isotropic case can be better
  understood. We therefore study the nearest-neighbor Heisenberg antiferromagnet
  on an anisotropic kagome lattice in a magnetic field. Such a system
  should be well described by weakly interacting spin chains, and we
  motivate a general form for the interaction by symmetry considerations
  and by perturbatively
  projecting out the inter-chain spins.  In the spin $1/2$ case, we
  find that the system exhibits a quantum phase transition from a
  ferrimagnetic ordered state to an XY ordered state as the field is
  increased. Finally, we discuss the appearance of magnetization
  plateaux in the ferrimagnetic phase.
\end{abstract}

\maketitle

\section{Introduction}

%
\begin{figure}[t]
\vskip0.5cm
\scalebox{0.4}{\includegraphics{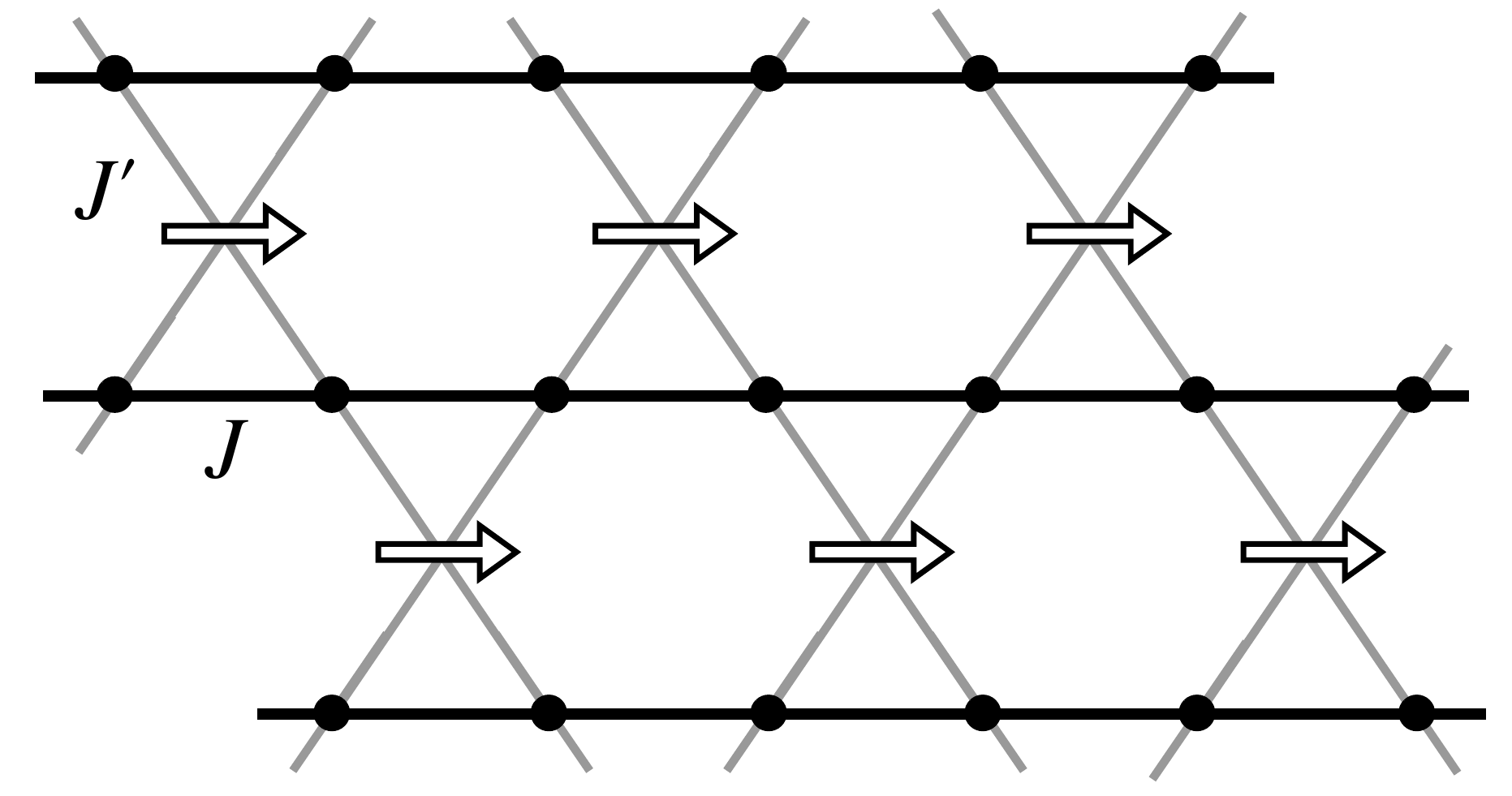}}
\caption{\label{kagome_lattice} The anisotropic kagome lattice with antiferromagnetic coupling $J$ between nearest-neighbor chain spins ($S$ spins) and coupling $J'$ between $S$ spins and inter-chain spins ($I$ spins). We make the approximation that the $I$ spins are fully polarized by the external magnetic field. }
\end{figure}
%

Frustrated antiferromagnets are of considerable interest because
frustrating interactions lead to strong fluctuations both of classical
(thermal) and quantum mechanical origin.  Unfortunately, even the
simplest model Hamiltonians for such materials are often not easily
analyzable by conventional theoretical techniques.  An outstanding
example is the nearest-neighbor quantum Heisenberg antiferromagnet on
the kagome lattice.  This structure is frustrated to a particularly
high degree, and the extensive classical ground state degeneracy may
be identified as a mechanism for strongly enhanced fluctuations. This
strongly limits the usefulness of the standard semi-classical spin
wave technique.  The behavior of the spin $S=1/2$ case in zero field
is particularly puzzling, with different numerical approaches yielding
contradictory and/or puzzling results~\cite{chalker-92,lecheminant-97,waldtmann-98,nikolic-03,cabra-05,misguich-07,singh-07}.  
Analytical studies have not
been any more illuminating as they neccessarily involve numerous
approximations that often lower the symmetry of the problem
substantially~\cite{huse-92,chubukov-92,sachdev-92,zeng-95,ran-06,ryu-07}.
However, it should be noted that recent experimental work on $\mathrm{ZnCu_3(OH)_6Cl_2}$ 
seems to favor a gapless, magnetically disordered ground state~\cite{keren-06,lee-07}.

One window into the behavior of frustrated antiferromagnets is their
{\sl magnetization process}, i.e.\ the curve of magnetization versus
applied field.  Variation of the magnetization over the full range
from zero to saturation opens another dimension of phase space in
which to explore the phase diagram of these systems, and perhaps find
physical and analytical insights.  A qualitative feature to be
understood is the occurrence of plateaux with quantized magnetization.
The structure of these plateaux reveal some aspects of the
correlations of the system.  For the low-spin kagome lattice, there
are suggestions of a plateau with $M=1/3M_s$ ($M_s$ is the saturation
magnetization).~\cite{hida-01,narumi-04,cabra-05, bergman-07,sen-07,sindzingre-07}
This is a very natural structure for the kagome
lattice, and can be understood as a state with two parallel and one
anti-parallel spins per triangle.  One experiment, however, suggests a
plateau with $M=1/2M_s$, which is not an obvious state~\cite{matsushita-06}.  

In this work, we consider an antiferromagnet on the \emph{anisotropic}
kagome lattice illustrated in fig.~\ref{kagome_lattice}, which may be
viewed as a set of spin chains (whose spins we call $S$ spins) coupled
together through intermediate spins ($I$ spins). The anisotropy could
stem from an inhomogeneity in the spin interactions, or from an actual
spatial anisotropy. A well known example of such a system is
volborthite, in which spin $1/2$ moments residing on copper atoms form
a kagome network that is anisotropic due to differing superexchange
bond angles.~\cite{hiroi-01}  Recently, there has been considerable
progress in developing techniques to analyze such systems of quantum
spin chains weakly coupled together by frustrating interactions.  Such
methods have the advantage that they can apply directly to low spin
and full Heisenberg ($SU(2)$) symmetry, and fully include the effects
of quantum fluctuations.  Here we  determine the physics of the
magnetization process in such a limit of the kagome lattice.  

We assume that the spins interact via a nearest neighbor Heisenberg
interaction, and we take the magnetic field to be applied in the $+z$
direction.  Taking the convention that lowercase indices only sum over
sites on the spin chains ($S$-spin sites) while uppercase indices only
sum over intermediate $I$-spin sites, we decompose the Hamiltonian as
\begin{eqnarray*}
{H} \  & = & {H}_0 + {H}^{\prime} \\ 
{H}_{0} & = & J \sum_{<ij>} \vec{S}_{i} \cdot \vec{S}_{j} + J^{\prime} \sum_{<Ai>} I^z_{A} S^z_{i} - h \sum_{i} S^{z}_{i} - h \sum_{A} I^{z}_{A} \\ 
{H}^{\prime} & = & \frac{ J^{\prime}}{2} \sum_{<Ai>} S^+_{i} I^-_{A} + S^-_{i} I^+_{A}\ .
\end{eqnarray*}
This decomposition is convenient in the limit we focus upon, namely
$J' \ll J$ and $J' \ll h$.  Note that we need not assume $h/J$ is
either large or small, and so can explore (almost) the full
magnetization process.  The region around $h=0$ is excluded however by
$J'\ll h$.  We leave this more challenging limit for future work.  We
have chosen to incorporate the $J'$ term coupling the z-components of
chain spins to inter-chain spins into $H_0$, which is convenient since
that term commutes with the large field ($h$) term.  This however
means that the expansions in powers of $H'$ and in $J'$ do not
strictly coincide, since the latter arise from two sources.

Because $J'$ is much smaller than the applied field, the
system may be described by a low-energy effective theory involving only
states in which the $I$ spins are fully polarized. 
In the limit $J'/J \rightarrow 0$, one expects that this low-energy effective theory will be a set of independent Luttinger liquids describing each chain (this will always be the case for
$S=1/2$ and will also hold for $S=1$ as long as the external field is larger than the zero-field gap to magnetic excitations). For very small $J'$, we continue to assume that the Luttinger liquid 
description is accurate, but there must also be weak inter-chain interactions. 
While the inter-chain interactions may be calculated, in principle, by symmetry considerations alone, we find it convenient to use an approach combining
perturbation theory  and symmetry arguments that allows the couplings to be estimated by hand.

For the case of a spin 1 system in a large enough field, we find that only one relevant interaction term emerges in 
an RG analysis of the general inter-chain coupling. Further analysis predicts that the resulting long-wavelength theory favors an XY antiferromagnetically
ordered phase at zero temperature. 

For the case of a spin $1/2$ system, however, an RG analysis of the inter-chain coupling reveals two relevant interaction terms. For large applied 
field, one interaction dominates the other in the long wavelength limit, leading to the same XY antiferromagnetically ordered phase as in the spin 1 case.
However, for small applied field, the other relevant interaction dominates, leading to a ferrimagnetically ordered phase. Finally, we reconcile
the discrete translational symmetry of the microscopic lattice with our continuum theory and show that magnetization plateaux may appear in the ferrimagnetic phase 
close to the transition. 
 
\section{Perturbative Calculation of Effective Hamiltonian}
  
Since we consider the limit $J'/J \ll 1$, the low energy states are those in which the
$I$ spins are completely polarized by the external field, and so we work
with an effective Hamiltonian projected into the space of such states.
We approximately calculate this effective Hamiltonian as follows.

\subsection{Elimination of inter-chain spins}

Let $P$ be the projection operator onto the subspace of all states
with the $I$ spins fully polarized, and let $Q=1-P$. Furthermore, for
a general state $\ket{\psi}$ let $\ket{\psi_\uparrow} = P \ket{\psi}$
and $\ket{\psi_Q} = Q \ket{\psi}$ so that $\ket{\psi} =
\ket{\psi_\uparrow} + \ket{\psi_Q}$.

Acting on the eigenvalue equation ${H} \ket{\psi} = E \ket{\psi}$ with $P$ and then with $Q$ gives the equations
\begin{eqnarray}
{H}_0 \ket{\psi_\uparrow} + P {H}' \ket{\psi_Q} & = &  E \ket{\psi_\uparrow} \label{Peq} \\
{H}' \ket{\psi_\uparrow} + {H}_0 \ket{\psi_Q} + Q {H}' \ket{\psi_Q} & = & E \ket{\psi_Q} \ . \label{Qeq}
\end{eqnarray}

Letting $R = (E-{H}_0)^{-1}$, (\ref{Qeq}) may be written as
\[
\ket{\psi_Q} = R {H}' \ket{\psi_\uparrow} + R Q {H}' \ket{\psi_Q}
\]
and then iterated to give
\begin{equation}
\ket{\psi_Q} = \frac{1}{1-R Q {H}'} R {H}' \ket{\psi_\uparrow} \ . \label{psiQ}
\end{equation}

Finally, upon replacing $\ket{\psi_Q}$ in (\ref{Peq}) by the expression in (\ref{psiQ}), we obtain an equation for $\ket{\psi_\uparrow}$ alone:
\begin{equation}
\left[ {H}_0 + P {H}' \frac{1}{1-R Q {H}'} R {H}' \right] \! \ket{\psi_\uparrow} \stackrel{\mbox{\tiny def}}{=} {H}_{\mbox{\tiny{eff}}} \ket{\psi_\uparrow} = \! E \ket{\psi_\uparrow}. \label{effeqn}
\end{equation}

Note, however, that Eq.(\ref{effeqn}) is not really a linear
eigenvalue equation, since $\heff$ itself depends on $E$ through $R$.

\subsection{Second Order Approximation}
We would like to use the assumption that $J' \ll J$ and $J' \ll h$ to simplify the expression for $\heff$ and find an approximate eigenvalue equation for $\ket{\psi_\uparrow}$. 
First, we make the generalization
\[
\mch \rightarrow \mch_\alpha = \mch_0 + \alpha\, \mch'
\]
in order to more conveniently keep track of powers of $\mch'$. Henceforth, take $\ket{\psi}$ to be the ground state of $\mch_\alpha$. The ground state energy
admits an expansion
\[
E = E_0 + E_1 + E_2 + \ldots
\]
where $E_n \propto \alpha^n$ and $E_0$ is the ground state energy of $\mch_0$.
$\heff$ may then be expanded as
\[
\heff = {H}_0 + {H}_2 + {O}(\alpha^3) , 
\]
with
\begin{equation}
  \label{eq:H2def}
  {H}_2 = P {H}' R_0 \mch' \ ,
\end{equation}
and $R_0 = (E_0 - \mch_0)^{-1}$. 

The second term in the expansion of $\heff$ may be simplified by use of the fact that we are restricting it to act only on states in the image of $P$. One finds
\begin{eqnarray*}
\lefteqn{{H}_2 \ket{\psi_\uparrow}  }\\
& = &\!\!  \frac{(\alpha J')^2 S}{2}\!\! \sum_{<Aik>} S^-_k \ \frac{1}{(E_0 - \mch_0) - h + J' \sum_{j \,\mbox{\scriptsize nn} A} S^z_j }\ S^+_i \ket{\psi_\uparrow}.
\end{eqnarray*}
Within the $P$ subspace, we may drop the ket and write
\begin{equation}
  \label{eq:loc}
  H_2 = \sum_A \mathcal{H}_2^{A},
\end{equation}
with
\begin{equation}
  \mathcal{H}_2^{A} = \frac{(\alpha J')^2 S}{2}\!\! \sum_{<ik;A>} S^-_k \ \frac{1}{(E_0 - \mch_0) - h + J' \sum_{j \,\mbox{\scriptsize nn} A} S^z_j }\ S^+_i, \label{eq:1}
\end{equation}
where $i,k$ are summed over the neighbors of $A$.  Note that, because of
the $J'$ dependence of the denominator in Eq.\eqref{eq:1}, $H_2^A$
contains non-trivial terms at all orders of $J'$ greater than or equal
to $O[(J')^2]$.  This is convenient because it allows us to obtain some
$O[(J')^4]$ contributions by an only $O(\alpha^2)$ calculation.

\section{Continuum Limit and Symmetry Considerations}

It is well-known that the low energy physics of an isolated
antiferromagnetic Heisenberg chain in a uniform applied field is
described by Luttinger liquid theory, whenever the field range is such
that the system remains gapless.  In the limit $J' \ll J,h$, which we
consider, all influences of the inter-chain spins upon the spin chains
indeed occur at low energies, and so can be considered in this
framework.  

Luttinger liquid theory consists of replacing the Hamiltonian of each
spin chain by that of a free boson field (and corrections that can be
analyzed perturbatively):
\begin{eqnarray}
\mch_0 & \rightarrow & \sum_n \mch^{\mbox{\tiny (LL)}}_n ,
\end{eqnarray}
where 
\begin{equation}
\mch_n^{\mbox{\tiny (LL)}} =  \frac{v}{2\pi} \sum_n \int_x  \left[\,  \frac{1}{g} \,(\partial_x \varphi_n)^2 + g\, (\partial_x \theta_n)^2 \right]
\end{equation}
is the Hamiltonian of a Luttinger liquid on chain $n$.  Here the
``spin velocity'' $v$ and the Luttinger parameter $g$ are known
functions of $h/J$ and $S$. The variables $\theta_n$ and $\varphi_n$
are dual boson fields living on the $n^{\mbox{\scriptsize th}}$ chain
and satisfying \mbox{$[\varphi_n(x), \theta_m(y)] = -i \pi \delta_{nm}
  \varphi(x-y)$}.

The physical interpretation of the phase fields is understood from
their relation to the microscopic spin operators:
\begin{eqnarray}
S^{\pm}_{j,n} & \sim & (-1)^j \,  e^{\pm i \theta_n(x_j)} [b_0 +b_1 \cos(2 \varphi_n(x_j) + Q x_j) ] \label{spm} \\
S^z_{j,n} & \sim & \mchain + \frac{1}{\pi} \partial_x \varphi_n(x_j) + a_1 \cos(2 \varphi_n(x_j) + Q x_j)\ . \label{sz}
\end{eqnarray}
The coefficients $a_1$, $b_0$ and $b_1$ depend on the chain
magnetization per site $\mchain$ and, for $S=1/2$, the wavevector $Q$
is given by $Q=2\pi (\mchain+1/2)$.

%
\begin{figure}[t]
\vskip0.5cm
\scalebox{0.35}{\includegraphics{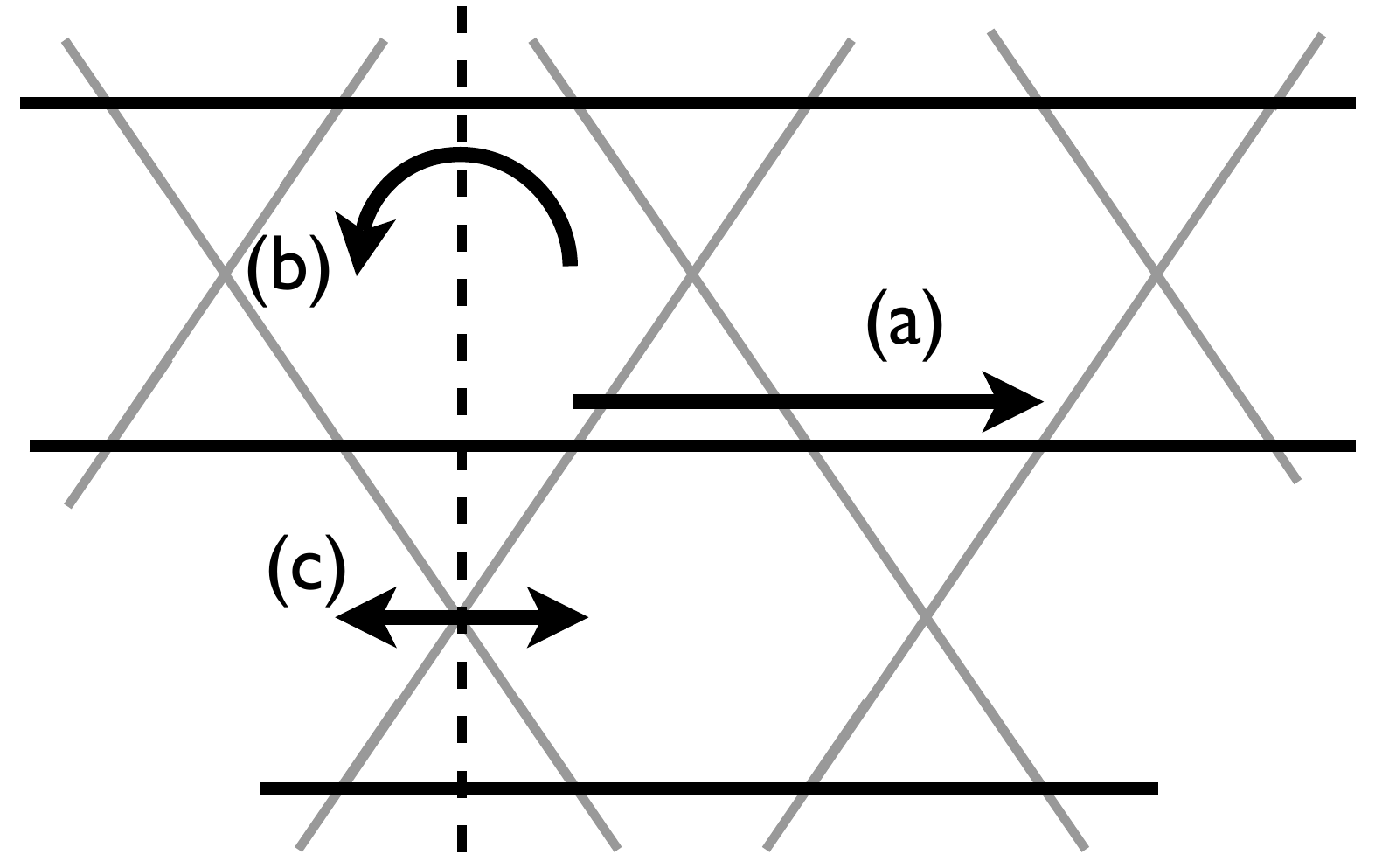}}
\caption{\label{symmetries} The generating symmetries of the space group \emph{cmm} of the 
anisotropic kagome lattice: (a) translations, (b) rotations by $\pi$ about the hexagon centers, and 
(c) reflections through a vertical line passing through the hexagon centers.  }
\end{figure}
%

From the above considerations, we expect that ${H}_2$ (and
higher order corrections) can also be expressed within the continuum
(conformal) field theory.  Formally, one may expand any local
Hamiltonian in terms of the scaling operators of the decoupled fixed
point theory.  Specifically, the density can be written as
\begin{equation}
  \label{eq:Hp}
  \mathcal{H}_2^A = \sum_\sigma \lambda_\sigma\,
  \mathcal{O}^{\sigma}_{n_A,n_A+1}(x_A),
\end{equation}
where $\mathcal{O}^{\sigma}_{n, n+1}(x)$ is a local symmetry allowed
operator involving degrees of freedom from the chains $n,n+1$
immediately above and below the site $A$, at horizontal position $x$,
with scaling dimension $\Delta_\sigma$.   This implies that the two
point functions of these operators obey
\begin{equation}
  \label{eq:5}
  \langle \mathcal{O}^\sigma_{n,n+1}(x)
  \mathcal{O}^\sigma_{n,n+1}(x')\rangle = \frac{C_\sigma}{|x-x'|^{2\Delta_\sigma}},
\end{equation}
with constants $C_\sigma$ that are dependent upon the convention for
normalizing the fields.  The expectation value is taken in the continuum
conformal field theory (CFT) describing the decoupled Heisenberg chains
in a field.  The two point function of two {\sl different}
operators $\sigma\neq\sigma'$ vanishes if the operators have different
symmetry (or more formally are descended from different primaries in the
CFT).  The sum over $\sigma$ may be thought of as being in order of
increasing scaling dimension.  Terms with large scaling dimension are
strongly irrelevant, and can therefore be neglected.

The first few terms in this expansion are strongly constrained by
symmetry.  Considering the full microscopic Hamiltonian, the
symmetries consist of the lattice space group (\emph{cmm} - see fig.\
\ref{symmetries}) and the $U(1)$ rotational symmetry of the spins
about the $z$ axis.  One finds
\begin{eqnarray}
\mathcal{H}_2^A & = & \frac{\mbox{}}{\mbox{}} \!\lambda_\perp
\cos[\theta_n-\theta_{n+1}] + \lambda_z  \cos[2\varphi_n-2\varphi_{n+1}]\label{eq:2}
\\
 & & \ \ \ \ \mbox{} + \lambda'_z\, \partial_x \varphi_n\, \partial_x
 \varphi_{n+1} \nonumber \\
 & & \ \ \ \ \mbox{} + \lambda'_\perp\, \partial_x \theta_n\,
 \partial_x \theta_{n+1} \cos[\theta_n-\theta_{n+1}]  + \ldots \!
 \frac{\mbox{}}{\mbox{}}  . \nonumber
\end{eqnarray}
Here we have suppressed the $A$ subscript on $n$ and $x$ on the right
hand side to keep the formula compact.

The second order Hamiltonian, $\mathcal{H}_2^A$, is actually further
constrained by an additional symmetry which one obtains only at this order.
Specifically, for each term in ${H}_2^A$ associated with an
inter-chain site $A$, the chains above and below this site can be {\sl
  independently} reflected about a vertical axis through site $A$.
This symmetry in fact requires that \mbox{$\lambda_\perp=0 + O[(J')^4]$}.

\subsection{General prescription for coefficients}
\label{sec:gener-prescr-coeff}

In this subsection, we demonstrate that the above effective continuum
Hamiltonian is uniquely determined from the microscopic
model, by showing how the coefficients ($\lambda_\sigma$) may be
obtained in principle from correlation functions of decoupled Heisenberg
chains.  In the following subsection, we will make some additional
simplifications in order to provide explicit expressions.  

In general, we can proceed by demanding equality between Eq.\eqref{eq:1} and
Eq.\eqref{eq:2} (Eq.\eqref{eq:Hp}).  The coefficients can then be
extracted by taking expectation values of these quantities with
microscopic operators whose continuum operator content is known, i.e.
\begin{equation}
  \label{eq:3}
  \hat{Q}_{j;n,n+1}^\sigma = c_\sigma \mathcal{O}^\sigma_{n,n+1}(x_j)+\cdots,
\end{equation}
where $\hat{Q}_{j;n,n+1}^\sigma$ is a microscopic expression composed of
lattice spin operators in the vicinity of site $j$ (horizontal position
$x'$) on chains $n,n+1$, and the omitted terms indicated by the ellipses
contain only operators of larger scaling dimension than
$\mathcal{O}^\sigma_{n,n+1}$.  For our problem, the needed
$\hat{Q}^\sigma$ operators and coefficients are
\[
\frac{1}{2} (S^+_{j,n} S^-_{j,n+1} + S^-_{j,n} S^+_{j,n+1}) = b_0^2 \cos[\theta_n-\theta_{n+1}] + \ldots 
\]
\[
S^z_{j,n} S^z_{j,n+1} =  \frac{a_1^2}{2}  \cos[2\varphi_n-2\varphi_{n+1}] + \frac{1}{\pi^2} \partial_x \varphi_n\, \partial_x \varphi_{n+1} + \ldots
\]
\begin{eqnarray}
\lefteqn{\frac{1}{2} (S^+_{j,n} + S^+_{j+1,n})( S^-_{j,n+1} + S^-_{j+1,n+1}) + \mbox{c.c.} = } \nonumber \\
& & \ \ \ \ \ \ \ \ \ \ \ = b_0^2 \partial_x \theta_n\, \partial_x
\theta_{n+1} \cos[\theta_n-\theta_{n+1}] + \ldots ,
\end{eqnarray}
where $a_1,b_0$ are the coefficients in Eqs.~(\ref{spm}, \ref{sz}).  

Knowing these, one has
\begin{eqnarray}
  \label{eq:4}
  \langle \hat{Q}_{j;n_A,n_A+1}^\sigma \mathcal{H}_2^A\rangle & = & c_\sigma
  \lambda_\sigma \langle
  \mathcal{O}^\sigma_{n_A,n_A+1}(x_j)\mathcal{O}^\sigma_{n_A,n_A+1}(x_A)\rangle \nonumber \\
  & = & \frac{c_\sigma \lambda_\sigma
    C_\sigma}{|x_j-x_A|^{2\Delta_\sigma}}+ \cdots.
\end{eqnarray}
The neglected terms decay faster with $|x_j-x_A|$, and can therefore be
distinguished from the dominant term above.  Thus by calculating the left
hand side in Eq.\eqref{eq:4}, and extracting its long-distance behavior,
one obtains the coefficient $\lambda_\sigma$, since $c_\sigma,C_\sigma$
are presumed known. 

\begin{widetext}
From this formulation, we can already deduce the order of the various
$\lambda_\sigma$ coefficients.  Consider first $\lambda'_\perp$.  We
require:
\begin{eqnarray}
  \label{eq:6}
&&  \left\langle \left[\frac{(S^+_{j,n_A} \!\!+\!\! S^+_{j+1,n_A})(
    S^-_{j,n_A+1} \!\!+\!\! S^-_{j+1,n_A+1})}{2} + \mbox{c.c.} \right]
  \mathcal{H}_2^A\right\rangle = \frac{b_0^2  C'_\perp
  \lambda'_\perp}{|x_j-x_A|^{2\Delta'_\perp}}.
\end{eqnarray}
The only obvious constraint to obtain a non-zero result for this
correlation function is that the sites $k,i$ in Eq.\eqref{eq:1} defining
$\mathcal{H}_2^A$ reside on neighboring chains $n_A,n_A+1$.  There are 4
such combinations included in the sum in Eq.\eqref{eq:1}.  Taking the
limit $J'\rightarrow 0$ in the resolvent denominator, we obtain
\begin{eqnarray}
  \label{eq:8}
&&   \!{(\alpha J')^2\, S\, }\bigg\langle
    \left[\frac{(S^+_{j,n_A} \!+\! S^+_{j+1,n_A})( 
    S^-_{j,n_A+1} \!+\! S^-_{j+1,n_A+1})}{2}\right]
\left[(S^-_{j_A,n_A} \!\!+\! S^-_{j_A+1,n_A})\frac{1}{E_0-H_0-h}( 
    S^+_{j_A,n_A+1} \!\!+\! S^+_{j_A+1,n_A+1}) \right]
  \bigg\rangle\nonumber \\ &&  = \frac{b_0^2  C'_\perp
  \lambda'_\perp}{|x_j-x_A|^{2\Delta'_\perp}}.
\end{eqnarray}
There is no reason for the correlation function on the left hand side to vanish.  To estimate
it, we note that the resolvent denominator is negative for all
eigenvalues and bounded below in magnitude by  $|E_0-H_0-h| >h$.  Hence
we estimate
\begin{equation}
  \label{eq:7}
  \lambda'_\perp \approxlt \frac{(\alpha J')^2}{h},
\end{equation}
up to $O(1)$ coefficients that can be smooth functions of $h/J$.

Consider now the remaining two non-zero coefficients, $\lambda_z,
\lambda'_z$.  Both may be obtained from the expectation value, which we denote $G$:
\begin{eqnarray}
  \label{eq:9}
  G& =& \left\langle S_{j,n_A}^z S_{j,n_A+1}^z \mathcal{H}_2^A\right\rangle = G_2 + G_3 + O[(J')^4] \ .
\end{eqnarray}
Since the number of particles in chains $n_A$ and $n_A+1$ must be
separately conserved, the sites $i,k$ in the sum in Eq.\eqref{eq:1} must
be on the same chain to obtain a non-zero result.  Hence we may write
\begin{eqnarray}
  \label{eq:10}
  G & \sim & (\alpha J')^2\, \left\langle S_{j,n_A}^z S^z_{j,n_A+1}
    \left[S_{j_A,n_A}^- + S_{j_A+1,n_A}^-\right]\frac{1}{(E_0 - \mch_0)
      - h + J' \sum_{j \,\mbox{\scriptsize nn} A} S^z_j }
    \left[S_{j_A,n_A}^+ + S_{j_A+1,n_A}^+\right]\right\rangle. 
\end{eqnarray}
First consider $G_2$, the second order in $J'$ contribution obtained by 
taking the $J'\rightarrow 0$ limit in the above resolvent. 
If one takes this limit, then the only remaining operator on chain
$n_A+1$ which is not translationally invariant is the single
$S^z_{j,n_A+1}$.  Applying a translation on chain $n_A+1$, then, one
sees that the $S^z_{j,n_A+1}$ can be replaced by the same operator at
any other site of the chain, i.e. $S^z_{j',n_A+1}$, with arbitrary
$j'$.  Hence, it can be replaced by average over all sites, i.e. the
uniform magnetization, which is a good quantum number and
non-fluctuation.  Upon making this substitution, one finds
\begin{eqnarray}
  \label{eq:11}
   G_2 & \sim & (\alpha J')^2 m_{\rm ch} \left\langle S_{j,n_A}^z 
    \left[S_{j_A,n_A}^- + S_{j_A+1,n_A}^-\right]\frac{1}{(E_0 - \mch_0)
      - h  }
    \left[S_{j_A,n_A}^+ + S_{j_A+1,n_A}^+\right]\right\rangle\\
  & \lesssim & - \frac{(\alpha J')^2}{h} m_{\rm ch}\left\langle S_{j,n_A}^z 
    \left[S_{j_A,n_A}^- + S_{j_A+1,n_A}^-\right]
    \left[S_{j_A,n_A}^+ + S_{j_A+1,n_A}^+\right]\right\rangle,\nonumber
\end{eqnarray}
where the $\lesssim$ is to be understood as indicating an inequality of
magnitude (not sign), using the same bound on the denominator as before.
The final line is straightforwardly analyzed from known results for
Heisenberg chains.  Because there is only a single field ($S_{j,n_A}^z$)
at position $x_j$, the long distance decay of $G_2$ contains
two pieces: an {\sl oscillatory} term decaying with the exponent
$\Delta_z$ and a non-oscillatory term decaying with the exponent
$\Delta'_z$.  The oscillatory term indicates a vanishing contribution
in the continuum limit.  The non-oscillating term has {\sl half} the
expected exponent, $2\Delta'_z$ to correspond to a contribution to
$\lambda'_z$.  It instead represents the generation of a term
proportional to $\partial_z\varphi_n$, which generates a small smooth
renormalization of the magnetization curve but is otherwise redundant.
Thus {\sl neither} $\lambda_z$ nor $\lambda'_z$ are generated at
$O[(J')^2]$.  

In order to calculate the leading contribution to $\lambda_z$ and $\lambda'_z$ then, one must 
calculate $G_3$ by expanding
the resolvent in Eq.(\ref{eq:10}) to first order in $J'$, giving
\begin{eqnarray}
\label{eq:11.5}
   G_3 & \sim & - J'\, (\alpha J')^2 \left\langle S_{j,n_A}^z S_{j,n_A+1}^z 
    \left[S_{j_A,n_A}^- + S_{j_A+1,n_A}^-\right]\frac{1}{(E_0 - \mch_0)
      - h  } \bigg{[}\sum_{k \,\mbox{\scriptsize nn} A} S^z_k \bigg{]} \frac{1}{(E_0 - \mch_0) - h}
    \left[S_{j_A,n_A}^+ + S_{j_A+1,n_A}^+\right]\right\rangle \nonumber \\ 
  & \lesssim & - \frac{J'\,(\alpha J')^2}{h^2} \left\langle S_{j,n_A}^z S_{j,n_A+1}^z
    \left[S_{j_A,n_A}^- + S_{j_A+1,n_A}^-\right] \bigg{[}\sum_{k \,\mbox{\scriptsize nn} A} S^z_k \bigg{]}
    \left[S_{j_A,n_A}^+ + S_{j_A+1,n_A}^+\right]\right\rangle \nonumber \\
   & \sim & \frac{a_1^2}{2} \frac{\lambda_z \, C_z}{|x_j - x_A|^{2 \Delta_z}} +  \frac{1}{\pi^2} \frac{\lambda'_z \,   C'_z}{|x_j - x_A|^{2 \Delta'_z}}
\end{eqnarray}
and so we find that $\lambda_z$ and $\lambda'_z$ are generally of order $(J')^3$.

\end{widetext}

\subsection{Explicit Calculation for Spin $1/2$ System}
\label{sec:explicit-calculation}

The above results provide a general formulation to calculate the
coefficients in the effective Hamiltonian, Eq.\eqref{eq:2}, from
correlation functions of a single Heisenberg chain in a field.  Because
of the complicated form of $\mathcal{H}^A_2$, however, even these
one-dimensional correlation functions are not simple to obtain
analytically, or extract from known results.  In this subsection,
therefore, we make some additional simplifying assumptions which allow
an explicit calculation of the effective Hamiltonian for the spin $1/2$ system.  
This calculation confirms the general structure obtained
above and allows for some semi-quantitative estimates.  

Consider again the the second order 
(in $\alpha$) contribution to effective Hamiltonian density
\begin{eqnarray*}
\lefteqn{\mathcal{H}_2^{A}=} \\
&\!\! &\!\! \left(\frac{\alpha J'}{2}\right)^2 \!\! \sum_{<ik;A>} S^-_k \ \frac{1}{(E_0 - \mch_0) - h + J' \sum_{j \,\mbox{\scriptsize nn} A} S^z_j }\ S^+_i.
\end{eqnarray*}
When the operator above acts within the low energy subspace, one would like to say that both the term proportional to $J'$ and the $(E_0 - \mch_0)$ term in the denominator are small, 
but it is not clear that $(E_0 - \mch_0)$ is small unless it acts \emph{directly} on the low energy states. So, we exchange the $S^+_i$ operator and the resolvent using spin $1/2$ anticommutation
relations to obtain
\begin{eqnarray*}
\lefteqn{\mathcal{H}_2^{A}=} \\
&\!\! &\!\! \left(\frac{\alpha J'}{2}\right)^2 \!\! \sum_{<ik;A>} S^-_k\ \left[\frac{\mbox{}}{\mbox{}} S^x_i\, R_{Ai;y} + i S^y_i\, R_{Ai;x} \right]\, .
\end{eqnarray*}
where
\begin{eqnarray*}
R_{Ai;\sigma} & = &  \left[ \frac{\mbox{}}{\mbox{}} (E_0 - \mch_0) -h + J' \sum_{j \,\mbox{\scriptsize nn} A} S^z_j   \right. \\
& & \left. \mbox{} -2h S^z_i  + 2J \sum_{j \,\mbox{\scriptsize nn} i} (S^z_i S^z_j + S^\sigma_i S^\sigma_j )\right]^{-1}\,.
\end{eqnarray*}

However, to be able to expand the $R_{Ai;\sigma}$ operators, we must make two rather unphysical assumptions. 
First, we take $J \ll h$ and second, we only let the magnetic field couple to $I$ spins.
This second assumption may not be too bad, however, since our eventual replacement of the chains of $S$ spins with Luttinger liquids takes the effect of the external field 
into account via the Luttinger parameter. 

After making the approximations just discussed and expanding, we finally obtain the following leading contributions:
\begin{eqnarray}
\mathcal{H}_2^A  & \approx & - \frac{(\alpha \Jp)^2}{4h} \sum_{<ik;A>} S^-_k S^+_i \nonumber \\
& & \mbox{} - \frac{(\alpha \Jp)^2 \Jp}{4h^2} \sum_{<ijk;A>} S^-_k S^+_i S^z_j  \nonumber \\
& & \mbox{} + \frac{(\alpha \Jp)^2 J}{4h^2} \sum_{<ik;A>} \sum_{j\, \mbox{\scriptsize nn}\, i} S^-_k S^+_i S^z_j -  S^-_k S^z_i S^+_j \ .\nonumber \\
& &  \label{2ndordercoupling}
\end{eqnarray}
In what follows, we set $\alpha$ back to $1$.

To connect this result to the continuum limit of the interchain coupling Eq.(\ref{eq:2}), we now proceed to bosonize the spin operators through the identifications in
Eq.(\ref{spm}) and Eq.(\ref{sz}). Keeping only contributions that involve products of operators from both chains, we find from the first term
\begin{eqnarray*}
\lefteqn{\sum_{<ik;A>} S^-_k S^+_i = } \\
& & (S^-_{j_A,n_A} + S^-_{j_A+1,n_A}) (S^+_{j_A,n_A+1}+S^+_{j_A+1,n_A+1} ) \\
& & \mbox{} + [n_A \leftrightarrow n_A+1] \\
& & = \,2 b_0^2 \,\, \partial_x \theta_n\, \partial_x \theta_{n+1} \cos[\theta_n-\theta_{n+1}] + \ldots
\end{eqnarray*}
which confirms that $\lambda'_\perp$ is of order $J'^2$, as we already argued by symmetry.

The second term in Eq.(\ref{2ndordercoupling}) must be treated with more care, however. Now, we only need to keep operators 
that conserve particle number on each chain, since we have already found the leading contribution to $\lambda'_\perp$ and since $\lambda_\perp = O[(J')^4]$ 
by symmetry. After also dropping operators that do not have contributions from both chains, we find
\begin{eqnarray*}
\lefteqn{\sum_{<ikj;A>} S^-_k S^+_i S^z_j =   (S^-_{j_A,n_A} \!\!+\! S^-_{j_A+1,n_A}) (S^+_{j_A,n_A} \!\!+\! S^+_{j_A+1,n_A} ) }\\
& & \times (S^z_{j_A,n_A+1} + S^z_{j_A+1,n_A+1}) + [n_A \leftrightarrow n_A+1] \\
& & = (S^z_{j_A,n_A+1} + S^z_{j_A+1,n_A+1}) \,(1 - S^z_{j_A,n_A} - S^z_{j_A+1,n_A} \\
& & \mbox{} +\! \{S^-_{j_A,n_A} S^+_{j_A+1,n_A} \!\!+\! S^+_{j_A,n_A} S^-_{j_A+1,n_A}\}) \!+\! [n_A\! \leftrightarrow \! n_A\!+\!1].
\end{eqnarray*}
It is the last operator in curly braces (the dimerization operator) that must be bosonized carefully. Since it involves products of operators at very short distances, all subleading 
bosonic operators omitted in Eq.(\ref{spm}) must be resummed, leading to the following continuum limit~\cite{hikihara-04}
\begin{equation}
(S^-_{j,n} S^+_{j+1,n} + S^+_{j,n} S^-_{j+1,n}) \sim c_0 + \frac{c'_0}{\pi} \partial_x \varphi_n - c_1\cos[2\varphi_n+Qx] .
\end{equation}
After using the above identity and collecting all relevant terms, we find that
\begin{eqnarray}
\lefteqn{\sum_{<ikj;A>} S^-_k S^+_i S^z_j = \frac{2}{\pi^2} (c'_0-4) \partial_x\varphi_{n_A} \partial_x\varphi_{n_A+1} }\\
& & \mbox{} - (2a_1^2 + a_1 c_1) (1+\cos{Q}) \cos[2\varphi_{n_A} - 2\varphi_{n_A+1}] + \ldots \nonumber
\end{eqnarray}
demonstrating that $\lambda_z$ and $\lambda'_z$ are both of order $J'^3$. The last term in Eq.(\ref{2ndordercoupling}) does not 
generate any relevant interactions in the bosonic theory.

To summarize, then, we have explicitly calculated the couplings in the effective Hamiltonian Eq.(\ref{eq:2}) for the spin $1/2$ system up to third 
order in $J'$ and the results confirm the predictions of our symmetry arguments (which apply to systems with any spin). The explicit
couplings are
\begin{eqnarray}
\lambda_\perp & = & 0 \nonumber \\
\lambda'_\perp & = & \frac{-J'^2\, b_0^2}{2h} \nonumber \\
\lambda_z & = & \frac{J'^3}{2 h^2}\,(2a_1^2+a_1 c_1) (1+\cos{Q}) \nonumber \\
\lambda'_z & = & \frac{J'^3}{\pi^2 h^2} (4-c'_0)\ . \label{bosonized_coupling}
\end{eqnarray}

\section{Analysis of the Continuum Model}

We have arrived, then, at a low energy effective theory describing the Heisenberg antiferromagnet on the anisotropic kagome 
lattice that takes the general form
\begin{eqnarray}
\lefteqn{\heff = } \nonumber \\
& & \sum_n \int_x  \left[ \ \ \frac{v}{2\pi} \frac{1}{g} (\partial_x \varphi_n)^2 + \frac{vg}{2\pi} (\partial_x \theta_n)^2  \right.  + \lambda'_z\, \partial_x \varphi_n\, \partial_x \varphi_{n+1} \nonumber \\
& &  \mbox{} +  \lambda_\perp  \cos[\theta_n-\theta_{n+1}] + \lambda_z  \cos[2\varphi_n-2\varphi_{n+1}] \nonumber \\
& & \left. \mbox{} + \lambda'_\perp\, \partial_x \theta_n\, \partial_x \theta_{n+1} \cos[\theta_n-\theta_{n+1}] + \ldots \begin{array}{l} \mbox{} \\ \mbox{} \end{array} \right]\, . \label{hamiltonian}
\end{eqnarray}

\subsection{RG Analysis}

To better understand the behavior of the above model at large length scales, we employ the momentum space RG procedure, 
lowering the cutoff from $\Lambda \rightarrow \Lambda - d\Lambda$ and defining $dl = d\Lambda/\Lambda$.
Including the marginal term  $\lambda'_z \, \partial_x \varphi_n\, \partial_x \varphi_{n+1}$ in the fixed-point action and expanding all 
quantities to first order in $\lambda'_z$  leads to the following tree-level flow equations for the other couplings (note that $\lambda'_z$ 
may be regarded as a constant independent of the scale of the effective theory):
\begin{eqnarray}
\frac{d\lambda'_\perp}{dl} & = & \left(- \frac{1}{2g} + \frac{\pi \lambda'_z}{4v} \right) \lambda'_\perp \nonumber \\
\frac{d\lambda_\perp}{dl} & = & \left(2-\frac{1}{2g} + \frac{\pi \lambda'_z}{4v} \right) \lambda_\perp + \frac{\pi \Lambda^2 \lambda'_z }{16 v} \lambda'_\perp \nonumber \\
\frac{d\lambda_z}{dl} & = & \left(2-2g- \frac{\pi g^2 \lambda'_z}{v} \right) \lambda_z \ . \label{flow_equations}
\end{eqnarray}

These flow equations imply that for $\lambda'_z \neq 0$ and $\lambda'_\perp \neq 0$, a finite value of $\lambda_\perp$
will be generated in the long-wavelength theory. 
Also, one can see that the $\theta$-interaction (with coupling $\lambda_\perp$) is relevant for $g \,{\scriptstyle\gtrsim} \, \frac{1}{4} - \frac{\pi \lambda'_z}{32 v}$ 
while the $\varphi$-interaction (with coupling $\lambda_z$) is
 relevant for $g \approxlt 1-\frac{\pi \lambda'_z}{2v}$.
 
 Now for the case of spin $1$ chain, F\'{a}th has determined numerically that $g$ is always greater than one (see F\'{a}th~\cite{fath-03}, fig.\ 5). In particular, $g$ is one
 at a chain magnetization per site of $\mchain=0$, reaches a maximum value of $g \approx 1.46$ at $\mchain \approx 0.36$ and approaches one again in the limit of full polarization. 
 So, for a spin 1 system described by (\ref{hamiltonian}), the $\theta$-interaction is always relevant, while the $\varphi$-interaction is always irrelevant.  
 
On the other hand, for a spin $1/2$ chain, the exact Bethe ansatz solution shows that $g$ increases nearly linearly from $1/2$ to $1$ as the chain magnetization per site increases from 
$\mchain=0$ to $\mchain=1/2$ (see Hikihara and Furusaki~\cite{hikihara-01}, fig.\ 2). So for a spin $1/2$ system described by (\ref{hamiltonian}), the $\varphi$-interaction and the $\theta$-interaction are nearly always relevant, and we must find a further criterion to determine which interaction dominates the behavior of the system. 
So, we solve the flow equations and determine which coupling 
constant becomes $\sim 1$ first as one continues to integrate out high momentum modes.

\subsection{Competing Interactions in the Spin 1/2 System}

We take the initial conditions given by Eq.(\ref{bosonized_coupling}). 
Then, defining
\begin{eqnarray*}
\beta'_\perp & =  & \left(-\frac{1}{2g} + \frac{\pi \lambda'_z}{4v} \right), \\
\beta_\perp & = & \left(2-\frac{1}{2g} + \frac{\pi \lambda'_z}{4v} \right)\\
\beta_z & = & \left(2-2g- \frac{\pi \lambda'_z g^2}{v} \right)
\end{eqnarray*}
the solutions to the flow equations for $\lambda_\perp$ and $\lambda_z$ are
\begin{eqnarray*} 
\lambda_\perp(l) & = & \frac{J'^5}{h^3} \frac{\Lambda^2 b_0^2}{64 \pi v} \,(4-c'_0) \left( e^{\beta_\perp l} - e^{\beta'_\perp l} \right) \\
 & \approx & \frac{J'^5}{h^3} \frac{\Lambda^2 b_0^2}{64 \pi v} \,(4-c'_0) \  e^{\beta_\perp l}  \\
\lambda_z(l) & = &  \frac{J'^3}{2 h^3}(a_1 c_1 + 2 a_1^2) (1+\cos{Q})\ e^{\beta_z l}\ .
\end{eqnarray*}

We can ignore the term $\sim e^{\beta'_\perp l}$ in $\lambda_\perp(l)$ since it rapidly goes to zero unless $g \approxgt 1/\lambda'_z$. 
Identifying $e^l$ as $L/a$, i.e.\ the ratio of the renormalized cutoff 
length scale to the bare lattice scale, we find that 
\begin{eqnarray*}
\lambda_\perp(l) & \sim & J'^5\ \left(\frac{L}{a}\right)^{\beta_\perp}   \\
\lambda_z(l) & \sim &J'^3\ \left(\frac{L}{a}\right)^{\beta_z}
\end{eqnarray*}

If we then define $L_\perp$ and $L_z$ as the length scales at which the $\theta$ and $\varphi$ couplings, respectively, become $\sim 1$, we find that
\[ 
\frac{L_z}{L_\perp} \sim (J')^{5/\beta_\perp - 3/\beta_z}
\]
The exponent is positive for $0< g < g_c$ and negative for $g_c < g \approxlt 1$, where $g_c \approx 0.636 - 0.635 \frac{\lambda'_z}{v} + \ldots$ 
This means that for $g < g_c$, $L_z \ll L_\perp$ and the $\varphi$ interaction is the dominant one.
For $g > g_c$, the $\theta$ interaction is dominant.

\subsection{Ordered Phases in the Gaussian Approximation}

\subsubsection{Ferrimagnetic Phase} 

First, consider a system described by the Hamiltonian
(\ref{hamiltonian}) in the limit where $g < g_c$, that is a spin $1/2$
system with magnetization per site between $m=0$ and $m \approx 1/5$
(on a scale where full polarization is $m=1/2$). In this limit, the
dominant relevant coupling is the $\varphi$-interaction with coupling
$\lambda_z$, so we drop all other couplings not in the fixed-point
action. The $\theta$ fields may then be integrated out to yield an
action entirely in terms of the $\varphi$ fields:
\begin{eqnarray}
S & \approx & \sum_n \int_\tau \int_x \ \frac{v}{2 \pi g} \left[ (\partial_x \varphi_n)^2 + \frac{1}{v^2} (\partial_{\tau} \varphi_n)^2 \right] \nonumber \\
 & & \!\!\!\!\!\!\!\!\mbox{} +  \lambda'_z \partial_x \varphi_n \partial_x \varphi_{n+1} + \lambda_z \cos[2\varphi_n - 2\varphi_{n+1}]\ . \label{theta_action}
\end{eqnarray}

Since the interaction term in the above theory is relevant,  the long-wavelength modes of the $\varphi$ fields will fluctuate very little from chain to chain if the coupling $\lambda_z$ is 
negative, while they will differ by $\pi/2$ from chain to chain if the coupling is positive. If $\lambda_z$ is positive, then, redefine the $\varphi$ fields by 
$\varphi_n(x) \rightarrow \varphi_n(x) + \frac{\pi}{2} n$. The coupling term for these new fields will now be negative. 

After the appropriate redefinition, fluctuations of the $\varphi$ fields from chain to chain will be suppressed at zero temperature, and so it is 
reasonable to expand the cosine term in the action to second order about zero. Upon performing a 
fourier transform in the chain index $n$, the resulting gaussian action is
\begin{eqnarray}
S &\!\! \approx &\!\! \frac{v}{2 \pi g} \int_{-\pi}^{\pi} \frac{dk}{2\pi} \int_{x,\tau} \left[ (1- \frac{2\pi g}{v} \lambda'_z \cos k)|\partial_x \varphi_k|^2 + \frac{1}{v^2} |\partial_{\tau} \varphi_k|^2 \right. \nonumber \\
 & & \left. +  m_\varphi^2(k)\, v^2|\varphi_k|^2 \right]\ \label{theta_gauss}
\end{eqnarray}
where the effective mass $m_\varphi$ is defined as 
\begin{equation}
m_\varphi^2(k) = \frac{8\pi g}{v^3}\, (1-\cos k)\, \lambda_z\ .
\end{equation}

A system described by the action (\ref{theta_gauss}) has excitations with a dispersion relation
\begin{equation}
\omega_{\varphi}^2(\vec{p},k) = (1- \frac{2\pi g}{v} \lambda'_z \cos k)\, p^2 v^2 + m_\varphi^2(k)\, v^4
\end{equation}
which vanishes as $\vec{p},\,k \rightarrow 0$, implying the existence of gapless excitations. 
Indeed, the action (\ref{theta_gauss}) is invariant under the transformation
\[
\varphi_{k=0}(x) \rightarrow \varphi_{k=0}(x) + \bar{\varphi} \ \ \ \mbox{($\bar{\varphi}$ constant)}
\]
because of the invariance of the system under translations along the chain or $x$-direction. 
The gapless excitations are the Goldstone modes associated with the breaking of this
symmetry.

%
\begin{figure}[t]
\vskip0.5cm
\scalebox{0.35}{\includegraphics{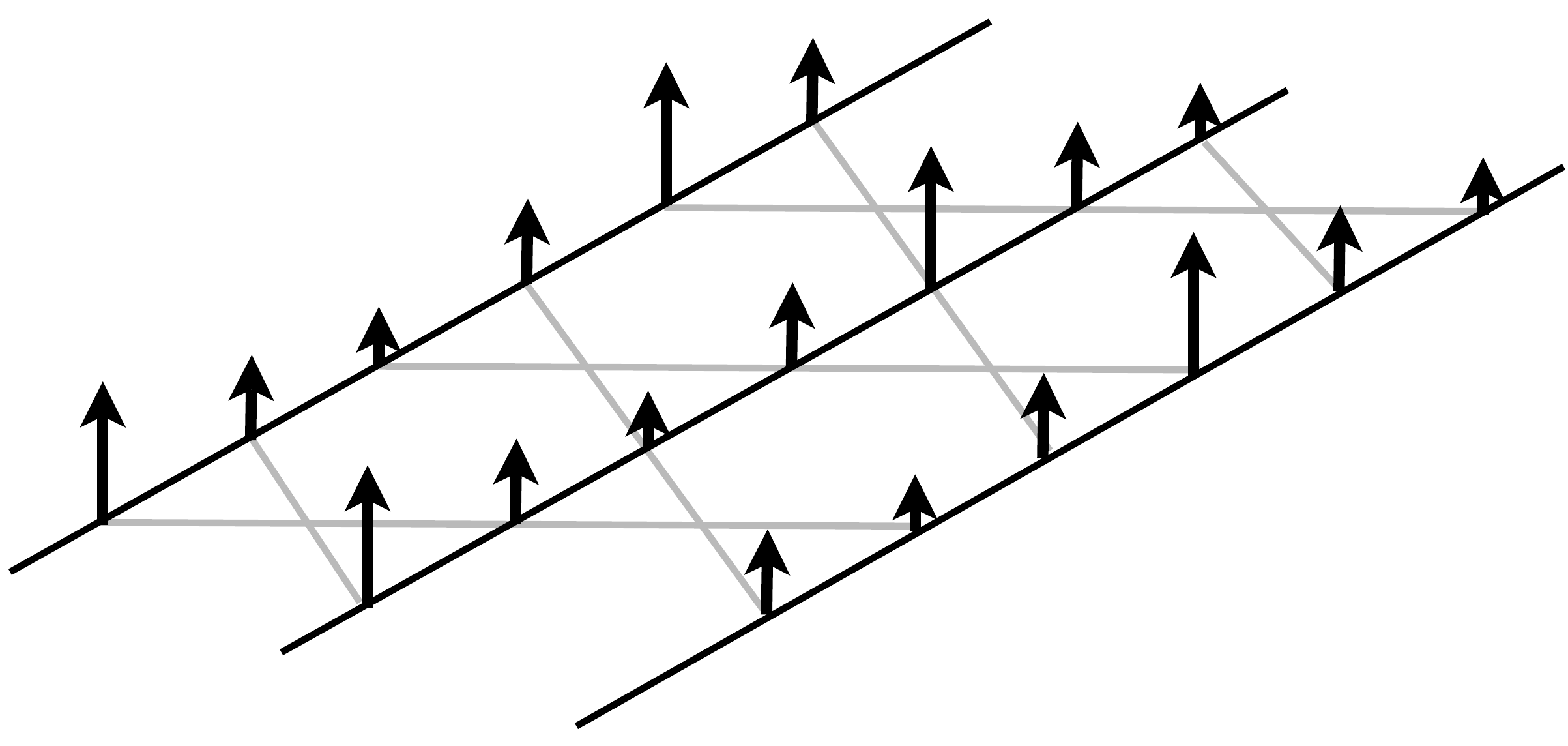}}
\caption{\label{ferrimagnetic_order} An example of a ferrimagnetically ordered ground state of the spin 1/2
system with magnetization $m \approxlt 1/5$ and $\lambda_z < 0$. }
\end{figure}
%
Restoring the original definition of the fields $\varphi_n(x)$ as appropriate, we see that
 for small enough $g$, the $\varphi_n$ fields will be ordered as $\langle \varphi_n(x) \rangle = \bar{\varphi} + \frac{\pi}{2} n$ if $\lambda_z$ is positive and 
 $\langle \varphi_n(x) \rangle = \bar{\varphi}$ if $\lambda_z$ is negative. 
Then, from Eq.(\ref{sz}) we see that the spins will be
ordered ferrimagnetically:
\begin{eqnarray} \label{ferri_order}
\lefteqn{\langle S_n^z(x)\rangle \sim} \nonumber \\
&& \left\{ \begin{array}{ll}
\mchain + \delta m\, (-1)^n \cos(2\bar{\varphi} + Q x) & \mbox{if $\lambda_z > 0$} \\
 \ \\
\mchain + \delta m\, \cos(2\bar{\varphi} + Q x) & \mbox{if $\lambda_z < 0$} \ .
\end{array} \right. \nonumber \\
& & \ 
\end{eqnarray}
(see fig.~\ref{ferrimagnetic_order} for an illustration).  The
amplitude of the ferrimagnetic spin density wave oscillation,$\delta m
\sim a_1 (J'/J)^{3g/(2-2g)}$ is small for small $J'$, due to the fact
that $S^z_n(x)$ is a continuum operator with all fluctuations on
length scales below $L_z$ integrated out. Since $g < 1$ in the spin
$1/2$ model, this factor will tend to suppress the oscillations in the
magnetization.

\subsubsection{XY Antiferromagnetic Phase}

Now, consider a system described by the Hamiltonian (\ref{hamiltonian}) but in the limit where $g > g_c$, namely a spin $1/2$ system 
with magnetization per site $m \approxgt 1/5$ or a spin $1$ system in a sufficiently large field (larger than both the Haldane gap and $J'$). 
In this limit, the dominant relevant 
coupling is the $\theta$-interaction with coupling $\lambda_\perp$, so we drop all other couplings not in the fixed-point action.

We proceed much as before, integrating out the $\varphi$ fields and redefining the $\theta$ fields  if $\lambda_\perp$ is 
positive. In this case, however, the appropriate redefinition is $\theta_n(x) \rightarrow \theta_n(x) + \pi n$. 
Once the fields are defined such that the cosine interaction makes them slowly varying at zero temperature, we expand the cosine to
obtain a gaussian theory once again, but with an effective mass 
\begin{equation}
m_\theta^2(k) = \frac{2\pi}{g v^3}\, (1-\cos k)\, \lambda_\perp
\end{equation}
and a dispersion
\begin{equation}
\omega_{\theta}^2(p,k) = (1- \frac{2\pi g}{v} \lambda'_z \cos k)\, [p^2 v^2 + m_\theta^2(k)\, v^4]\ .
\end{equation}

%
\begin{figure}[t]
\vskip0.5cm
\scalebox{0.35}{\includegraphics{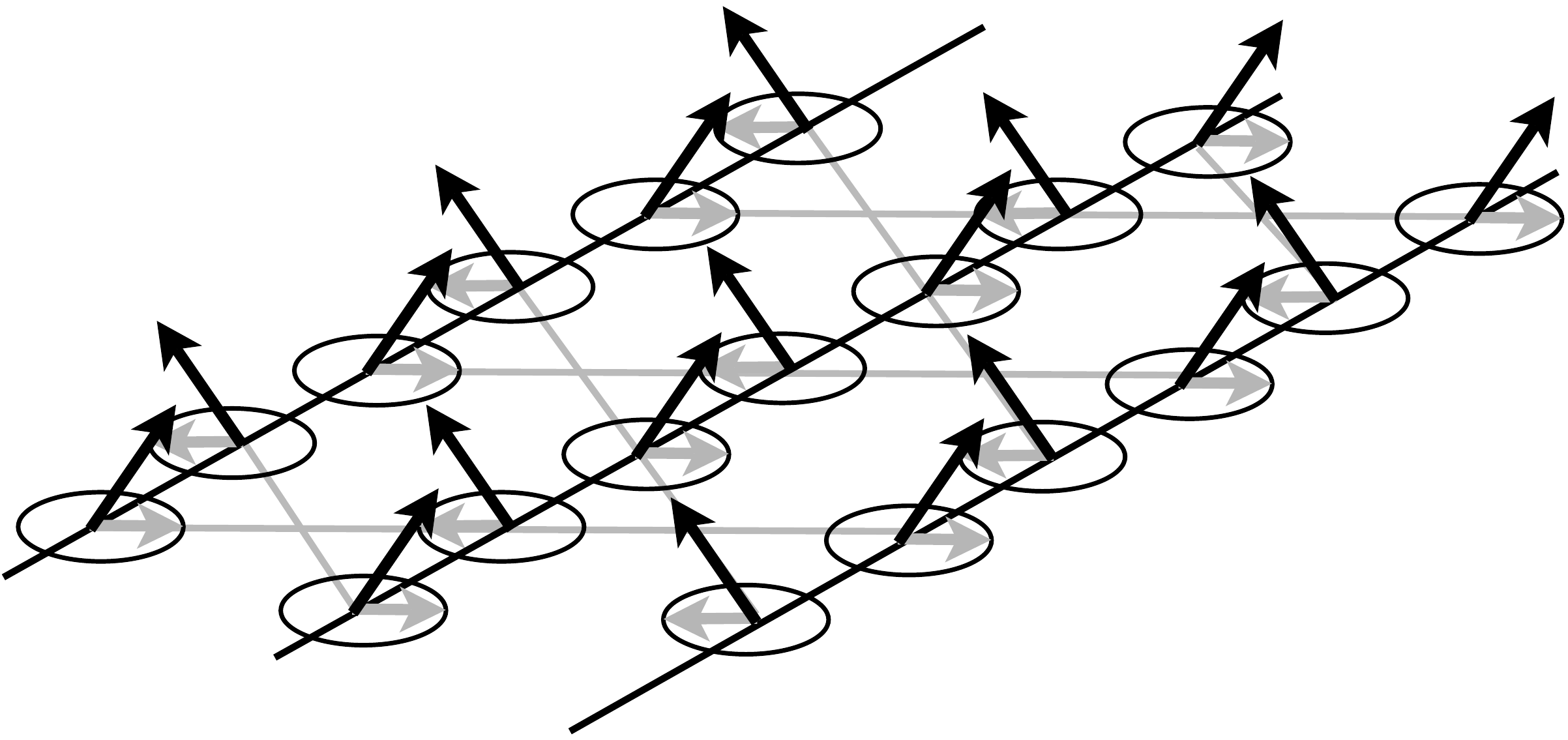}}
\caption{\label{xy_order} The XY antiferromagnetically ordered ground state that occurs for both the spin 1/2 and 
spin 1 systems in a large enough field ($\lambda_\perp < 0$ case shown). }
\end{figure}
%

The gapless modes in this limit are the Goldstone modes associated with the breaking of the 
shift symmetry 
\[
\theta_{k=0}(x) \rightarrow \theta_{k=0}(x) + \bar{\theta}\ \ \ \mbox{($\bar{\theta}$ constant)}
\]
associated with the global rotational symmetry of all spins about the magnetic field direction. 

Restoring the original fields, we expect that, for large enough $g$, $\langle\theta_n(x)\rangle = \bar{\theta} + \pi n$ if $\lambda_\perp > 0$ or 
$\langle\theta_n(x)\rangle = \bar{\theta}$ if $\lambda_\perp < 0$.
From the identification (\ref{spm}), we therefore expect an XY antiferromagnetic 
ordering of the spins:
\begin{equation}
\langle S^+_{j,n} \rangle \sim  \left\{ \begin{array}{ll}
(-1)^{j+n}\, m_s e^{i\bar{\theta}} & \mbox{if $\lambda_\perp > 0$} \\
(-1)^j\, m_s e^{i\bar{\theta}} & \mbox{if $\lambda_\perp < 0$}\ . 
\end{array} \right. ,
\end{equation}
where $m_s \sim b_0 (J'/J)^{5/(8g-2)}$ is the staggered magnetization,
which is reduced by fluctuations below the scale $L_\perp$.  See
fig.~\ref{xy_order} for an illustration.

\subsection{Lattice Effects in the Spin $1/2$ Ferrimagnetic Phase}

\subsubsection{Commensurate Ferrimagnetic Order}

For a spin $1/2$ system with $g < g_c$, the ordering of the zero mode of the $\varphi$ field corresponds to the breaking of the translational symmetry of the system in the $x$ direction
(along the chains). However, this symmetry is not actually continuous in the microscopic theory, but reduces to the discrete translational symmetry of the
lattice.

In order to reconcile the continuum and microscopic theories then, we observe that the Hamiltonian (\ref{hamiltonian}) may also contain terms of the 
form $\cos[2 p\, \varphi_n(x)]$ for certain values of $p$ that must be determined by symmetry.

Under translation by one lattice spacing along the chain direction, the $\varphi$ fields transform as
\begin{equation}
\varphi_n(x) \rightarrow \varphi_n(x-1) - Q/2 \ .
\end{equation}
So, for a term of the form $\cos[2 p\, \varphi_n(x)]$ to respect the translational symmetry of the system,
it must be that $p\,Q = 2 \pi k$, where $k$ is an integer. Moreover, the action should be invariant under $\varphi_n(x) \rightarrow \varphi_n(x) + \pi$, so $k$ 
must be chosen to make $p$ an integer.

Recalling that $Q = 2\pi (\mchain + 1/2)$, both requirements imply that only the set $\{p_j\}$ are allowed, where 
\begin{equation}
p_j = \frac{k_j}{\mchain+\frac{1}{2}}  \label{allowed_ps}
\end{equation}
and $k_j$ is the $j$th positive integer (in increasing order) such that $p_j$ is an integer. We may identify
\begin{equation}
\bar{\lambda} = \frac{1}{\mchain+\frac{1}{2}} = \frac{2\pi}{Q}
\end{equation}
as the wavelength of the (continuum) ferrimagnetic wave (\ref{ferri_order}).
But, for a general magnetization, the wavelength $\bar{\lambda}$ will not be an integer multiple of the lattice spacing. Instead, the ferrimagnetic wave above, when 
restricted to values of $x$ that coincide with the lattice, will appear to be periodic with an effective wavelength $\lambda_{\mbox{\tiny eff}} = k_0 \bar{\lambda} = p_0$. 

So then, given that for a fixed magnetization the corresponding term $\cos[2 p_0\, \varphi_n]$ will appear in the Hamiltonian, it
will bring the discrete translational symmetry of the lattice to bear on the ordered phase by restricting the 
values of $\langle \varphi_n(x)\rangle = \bar{\varphi}$ to coincide with its minima. 
For a negative coefficient of  $\cos[2 p_0\, \varphi_n]$, this will force the ferrimagnetic oscillations to be symmetric about a particular lattice site.
On the other hand, if the coefficient is positive the oscillations will be symmetric about the point on the lattice 
between two particular neighboring sites. 

\subsubsection{Magnetization Plateaux}

In addition to pinning the ferrimagnetic order to the lattice, the symmetry-allowed $\cos[2 p\, \varphi_n(x)]$ terms in the Hamiltonian may lead to the appearance of 
plateaux in the magnetization curve of the spin 1/2 system.

For general (but small) values of the external field, the ground state of the spin $1/2$ system will be, in bosonic terms,
a state in which the $\varphi_n(x)$ fields on each chain fluctuate about some constant average value as we discussed in detail above 
(we take the coupling $\lambda_z$ to be negative in this section for simplicity - the other case follows straightforwardly). 
In such states, the chain magnetization per site smoothly oscillates as a function of $x$ about its average (to zeroth order in $J'$) value $\mchain$. 
Note that $\mchain$ is simply the magnetization of a decoupled Heisenberg chain and as such, is a monotonially increasing function of the external magnetic field.\cite{hikihara-01}
 
%
\begin{figure}[t]
\vskip0.5cm
\scalebox{0.35}{\includegraphics{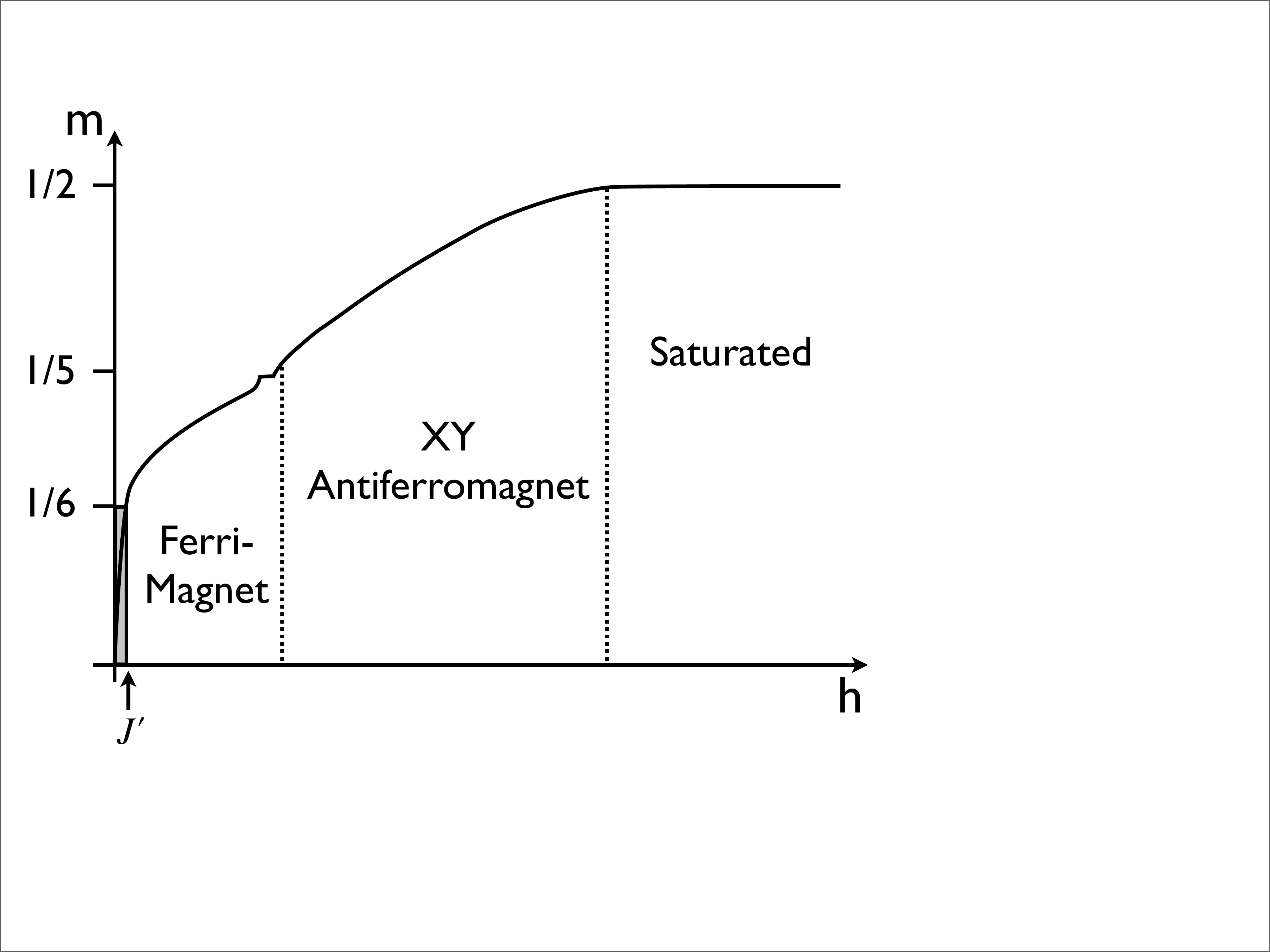}}
\caption{\label{phase_diagram} Schematic graph of the magnetization curve and phases of the spin-1/2 system. 
Note the hypothetical plateau just below the crossover point. 
The shaded region indicates the parameter space
where our model does not apply.   }
\end{figure}
%

Now consider some particular magnetization $m^P$ (which corresponds to a chain magnetization $\mchain^P = 3m^P/2 -1/4$) and first assume that the  
external field has been tuned to a value $h^P$ such that $\mchain(h^P) = \mchain^P$. In such an external field, the most important
symmetry allowed term of the type $\cos[2 p\, \varphi_n(x)]$ that may be added to the Hamiltonian is the one with the smallest 
value of $p$ given by Eq.(\ref{allowed_ps}), namely 
\begin{equation}
p_0 = \frac{k_0}{\mchain^P+\frac{1}{2}} , \label{p0plateau}
\end{equation}
where $k_0$ is defined as the smallest positive integer such that $p_0$ is an integer.

To illustrate, then, how a plateau may arise at this magnetization, we consider the Hamiltonian 
\begin{eqnarray}
\heff \!\!\! & = \!\!  &\! \sum_n\! \int_x \frac{v}{2 \pi g} \left[ (\partial_x \varphi_n)^2 + \frac{1}{v^2} (\partial_{\tau} \varphi_n)^2 \right] \!+\! \lambda'_z \partial_x \varphi_n \partial_x \varphi_{n+1}  \nonumber \\
& &  \!\!\!\!\mbox{} - |\lambda_{z0}| \cos[2 (\varphi_n - \varphi_{n+1})] - \eta_0 \cos[2 p_0 \varphi_n + \delta x] + \ldots \nonumber \label{plateau_action} \nonumber \\
& & \ 
\end{eqnarray}
and no longer think of the external field as fixed, but as varying in some small range about $h^P$. Since the coefficient $\delta$ is zero for $h = h^P$
by the definition of $p_0$,  
symmetry considerations show that it must be given in general by
\begin{equation}
\delta(h) = 2\pi p_0 [\mchain(h) - \mchain^P] \ .
\end{equation}

Now, if we integrate out the short wavelength modes of the system in an iterative RG procedure 
(but \emph{without} rescaling the high momentum cutoff), 
the effective coupling $\lambda_z$ will vary as
\begin{equation}
\lambda_z[L] \sim \lambda_{z0} \, (L)^{-2g} 
\end{equation}
\\
until we reach a length scale $L_z$ defined such that $\lambda_z[L_z] \sim 1/(L_z)^2$ (i.e.\ this is the 
scale such that $\lambda_z$ would be $O(1)$ if we were rescaling at every RG step). Therefore, we find that 
\begin{equation}
L_z \sim (\lambda_{z0})^{-1/(2-2g)}\ .
\end{equation}
At this scale, the chains will be strongly coupled and the $\varphi$ fields will vary slowly from chain to chain. 

Then, since the $\varphi$ fields vary slowly from chain to chain, it is reasonable to take a continuum limit  $\varphi_n(x) \rightarrow \varphi(x,y)/\sqrt{d}$ 
(where $d$ is the chain spacing). Within such an effective $(2+1)$-dimensional theory, the long-wavelength fluctuations 
of the field $\varphi(x,y)$ are bounded and so the term in the action $\eta_0 \cos[2 p_0 \varphi_n]$, which 
may have been an irrelevant perturbation to a decoupled chain is now a relevant perturbation, regardless of the value of $p_0$. 

Now, since the scaling dimension of the $\cos[2 p_0 \varphi]$ term is $p_0^2\, g$, the effective coupling $\eta$ becomes 
\begin{equation}
\eta(L_z) \sim \eta_0\cdot (L_z)^{-p_0^2g} \sim \eta_0 \cdot (\lambda_{z0})^{\frac{p_0^2 g}{2-2g}}
\end{equation}
as the momentum cutoff is lowered to $1/L_z$ under the RG procedure. Since $\lambda_{z0}$ is small (generally $\sim J'^3$) and $g<1$ for the spin $1/2$
system in its ferrimagnetic phase, this renormalized value of $\eta$ will
decrease rapidly as a function of $p_0$. 

Finally, then, we determine if a plateau develops around the magnetization $m^P$ by considering the competition between the kinetic energy terms 
in $\heff$ and the term with coupling $\eta$. The latter term is minimized if $\langle \varphi_n(x) \rangle = \frac{-\delta(h)}{2 p_0}\, x$
(that is, if $\langle S^z_n(x) \rangle = \mchain^P$ regardless of the external field - a plateau).
However, such a field configuration costs some kinetic energy and the resulting total change in the linear energy density is 
\begin{equation}
\Delta E_n(x) \sim \frac{v}{2\pi g} \bigg( \frac{\delta(h)}{2 p_0} \bigg)^2 - \eta = \frac{v \, [\delta(h)]^2}{8 \pi g p_0^2} - \eta \ .
\end{equation}
Therefore, we see that this arrangment remains favorable (i.e.\ the plateau exists) as long as 
\begin{eqnarray}
|\delta(h)| & < &  \sqrt{\eta} \ 2 p_0 \sqrt{\frac{2\pi g}{v}} \\
\implies |\mchain(h) - \mchain^P| & \approxlt & (J')^{3p_0^2 g/4(1-g)} \ .\nonumber \label{plateau_width}
\end{eqnarray}
A significant plateau will only appear, then, for a relatively small $p_0$, which is itself determined by $\mchain^P$ through~(\ref{p0plateau}).

We therefore expect the widest plateau, if one appears at all, to appear just below the crossover point from the ferrimagnetic to the 
XY antiferromagnetic phase in the spin $1/2$ system (see fig.~\ref{phase_diagram}). Furthermore, we may estimate the dependence of the plateau width on $J'$ from our
earlier estimate that $g_c \sim 0.636$ and from the fact that the smallest $p_0$ allowed by Eq.(\ref{p0plateau}) near the crossover is about 9. 
Using these values in Eq.(\ref{plateau_width}), the power of $J'$ is $\sim 100$, so any plateau near the crossover will be narrow indeed.

\begin{acknowledgements}
This work was supported by the Packard Foundation and the National Science Foundation through grant DMR04-57440.
\end{acknowledgements}





\bibliography{anisotropic_kagome2}

\end{document}